\documentclass[fleqn,twoside]{article}
\usepackage{espcrc2}

\usepackage{graphicx}
\usepackage{axodraw}

\usepackage[figuresright]{rotating}

\newcommand{\AmS}{{\protect\the\textfont2
  A\kern-.1667em\lower.5ex\hbox{M}\kern-.125emS}}

\newcommand{\PRD}[3]{{\it Phys.\ Rev.\ }     {\bf D#1}~{(#2)}, {#3}}
\newcommand{\NP}[3]{{\it Nucl.\ Phys.\ }         {\bf #1}~{(#2)}, {#3}}
\newcommand{\CMP}[3]{{\it Commun.\  Math.\  Phys.\  }  {\bf #1}~{(#2)}, {#3}}

\title{Effects of Quenching in $\Delta I=1/2$ Kaon Decays\thanks{SHEP 02-19}}

\author{
  C.-J.~D.~Lin\address[SO]{Dept. of Physics and Astronomy, Univ. of Southampton,
    Southampton, SO17 1BJ, UK}, 
  G.~Martinelli\address[RM]{Dip. di Fisica, Univ. di Roma ``La Sapienza'' and INFN, \\Sez. di Roma, P.le A. Moro 2, I-00185 Rome, Italy}, 
  E.~Pallante\address{SISSA, Via Beirut 2-4, 34013, Trieste, Italy}, 
  C.~T.~Sachrajda\addressmark[SO]$^{,}$\address{Theory Division, CERN, CH-1211 Geneva 23, Switzerland} and
  G.~Villadoro\addressmark[RM]\thanks{presenter at the conference.}}
       
\begin{document}

\begin{abstract}
We present the inconsistencies which arise in quenched QCD in $\Delta I=1/2$ 
non-leptonic kaon decays using chiral perturbation theory ($\chi$PT) to one 
loop. In particular we discuss how the 
lack of unitarity of the quenched theory invalidates the usual methods for
the extraction of matrix elements from correlation functions. 
\vspace{1pc}
\end{abstract}

\maketitle

\section{INTRODUCTION}
\label{sec:intro}
While different techniques have been developed in order to extract
$\Delta I=3/2$ $K\rightarrow\pi\pi$ matrix elements from the lattice 
(\cite{02Boucaud,02Lin}) and finite volume corrections can be controlled 
even in the quenched approximation (at least to one loop in $\chi$PT)  
(\cite{02Lin,01Lellouch,01Lin,92Bernard,97Golterman}), this is not true
for the $\Delta I=1/2$ channel. In this case, 
in addition to the presence of large final state interactions (FSI), 
there are manifest inconsistencies due to quenching.
The reason why the $\Delta I=3/2$ case is different will be explained in
sec.~\ref{sec:scatt}, while the explicit calculation of the scalar 
form factor (which shares the same quenching artifacts as $K\to\pi\pi$, 
$\Delta I=1/2$) is in ref.~\cite{02scalar}. 
Some of these problems in the $\chi$PT framework were first 
discovered in 
refs.~\cite{92Bernard,95Bernard,00Golterman}, our aim is to exhibit the 
full set of the quenching effects and to understand their origin.
By computing matrix elements involving isoscalar $\pi\pi$ scattering 
contributions (e.g. $K\to\pi\pi$ or the scalar form factor) 
using quenched $\chi$PT ($Q\chi PT$), in finite
and infinite volume, with the insertion of different $\Delta I=1/2$ 
weak operators we observe that:

1) The strong phase shift is no longer universal since it depends on 
the choice of the weak operator.

2) The L\"uscher quantization condition and the
Lellouch-L\"uscher ($LL$) finite-volume correction factor for matrix
elements are no longer valid.

3) The standard procedure for extracting matrix elements
from time-independent ratios of finite-volume
euclidean correlation functions (CF) fails, 
since the time dependence of the CF depends on the choice of the two-pion sink.

4) The $\eta'$ double-pole produces new polynomial terms in 
$t$ (time) and $L$ (box size) which cause practical problems in the extraction 
of matrix elements.

\section{FULL (UNQUENCHED) QCD}
\label{sec:cf}
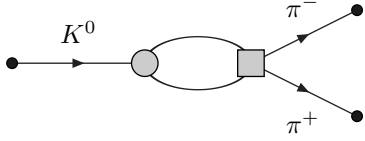
\begin{figure}[t]
\begin{center}
\begin{picture}(0,50)(70,80)
\ArrowLine(0,100)(50,100) \ArrowLine(90,100)(130,120)
\ArrowLine(90,100)(130,80) \Oval(70,100)(10,20)(0)
\GCirc(0,100){2}{0.1}\GCirc(130,120){2}{0.1}
\GCirc(130,80){2}{0.1} \Text(25,108)[b]{$K^0$}
\Text(110,118)[b]{$\pi^-$}\Text(110,82)[t]{$\pi^+$}
\GCirc(50,100){5}{0.8} \GBoxc(90,100)(10,10){0.8}
\end{picture}
\end{center}
\caption{\label{fig:scatt} \it The one-loop diagram which contains power corrections in the volume.}
\end{figure}

The unquenched $K\to\pi\pi$ matrix element can be extracted by 
studying the time behaviour of a suitable four-point euclidean 
CF in finite volumes:
\begin{eqnarray}
\sum_{\Omega_{\vec q}} \langle 0 | \pi^-_{\vec q}(t)\pi^+_{-\vec q}(t)
H_W(0)K^\dagger_{\vec 0}(t_K)|0\rangle =  \nonumber \\
\sum_{\Omega_{\vec q}} \langle 0 | \pi^-_{\vec q}(0)\pi^+_{-\vec q}(0)|W\rangle
\langle W|H_W(0)|K\rangle \times \nonumber \\
\times\langle K |K^\dagger_{\vec 0}(0)|0\rangle e^{-m_K|t_K|-W|t|} +\ldots
 \, , \label{eq:cf}
\end{eqnarray} 
where $\phi_{\vec q}(t)=\int d^3 {\vec x}\phi({\vec x},t)e^{i \vec{q}
\cdot\vec{x}}$, 
$|W\rangle$ is a two-pion state with total energy $W$, 
$\Omega_{\vec q}=\{{\vec q}:|{\vec q}|\, fixed\}$ and the large time limit 
has been taken while isolating the contribution
with time behaviour $e^{-W|t|}$. 
In order to extract the matrix element $\langle W|H_W(0)|K\rangle$ one has 
to divide by the factors corresponding to the kaon
source and the two-pion sink extracted from the corresponding 
CF~\cite{02Lin}.
The validity of eq.~(\ref{eq:cf}) and Watson's theorem rely on
unitarity, and it is the absence of unitarity in the quenched
theory which is responsible for most of the difficulties listed at
the end of sec.~\ref{sec:intro}.
The calculation in infinite volume is
straightforward and we will not discuss it further, except to
point out that of the one-loop diagrams, it is only that in fig.~(\ref{fig:scatt})
which can have physical intermediate states and hence an imaginary
part (and $1/L^n$ corrections in finite-volumes).

\subsection{Finite volume}
Evaluating eq.~(\ref{eq:cf}) to one loop in $\chi$PT on a finite volume (FV)
one obtains, up to exponentially small terms in the volume,
\begin{eqnarray}
\label{full}
&&\frac{e^{-m_K |t_K|}}{2m_K}\frac{e^{-2E|t|}}{(2E)^2}A_{\infty}(E)[1+T] 
\, , \\
T&=&-\Delta W t +\frac{\partial A_{\infty}}{\partial W}\Delta W 
+\frac{b_s}{L^3}+\nonumber \\
&&+\frac{c_0 z(0)}{(EL)^3}+\frac{c_1 z(1)}{EL} \nonumber \, ,
\end{eqnarray}
where $A_{\infty}$ is the corresponding infinite volume result, 
$T$ parameterizes 
FV corrections, $\Delta W$ is the energy shift ($\propto 1/L^3$), 
$E^2={\vec q}^{\,2}+m_\pi^2$ and $z(i)$ are defined for example in
\cite{02Lin,01Lellouch}.
The first two terms in  $T$ shift the two-pion energy in the time 
exponential and in the argument of $A_{\infty}$ respectively, 
the third term depends on the sink used to 
annihilate the two pions (it is cancelled when the matrix element is extracted
\cite{02Lin}) and finally the last two terms represent the chiral expansion 
of the $LL$ finite volume factor to this order. 
The validity of eq.~(\ref{full}) relies on the unitarity of the theory.

\section{QUENCHED QCD}
\label{sec:scatt}
We now consider the determination of the matrix elements in
quenched QCD, restricting our attention to the diagram in fig.~(\ref{fig:scatt}),
which is the one leading to anomalous effects. In particular it
receives two unphysical contributions: (i) ghosts which cancel
internal quark loops, (ii) $\eta'$ with double-pole insertions~(\footnote{ 
Both these contributions do not appear to one loop in the $\Delta I=3/2$ case 
because of total isospin conservation.}). 
They lead to a number of inconsistencies:

(i) The ghosts are a manifestation of the lack of unitarity.
Phase shifts, energy shifts and finite volume corrections are no
longer universal, the factor corresponding to the sink cannot be
removed and the extraction procedure in eq.~(\ref{eq:cf}) cannot be applied.

(ii) $\eta'$ double-poles lead to a more severe and unphysical
infrared structure. In infinite volume they
appear as divergences in the chiral limit 
(the so-called ``quenched chiral logs'', see for example \cite{92Bernard}) 
and in the singular limit at threshold \cite{02scalar,98Colangelo}. 
In finite volumes they produce terms with an anomalous
behaviour in time and volume (see below and 
\cite{02scalar,95Bernard,00Golterman}).

\subsection{The scalar form factor}
To study these effects consider the time-like scalar form 
factor of a scalar and isoscalar operator $S$, 
which shows all the effects of $K\to\pi\pi$, $\Delta I=1/2$
 but has a simpler chiral structure. The corresponding CF 
at one loop in Q$\chi$PT can be written as follows:
\begin{eqnarray}
\label{scalarff}
\langle 0 |\pi^+_{\vec q}(t) \pi^-_{-\vec q}(t) S(0) |0\rangle=
\frac{e^{-2E|t|}}{(2E)^2}A_{\infty}[1+T]\,,\quad \mbox{}  \\
T=\frac{a_1 t +a_2 t^2+a_3 t^3}{(EL)^3}+\frac{b_0 +b_1 t +b_2 t^2}{(EL)^3}
+\quad \mbox{}
\nonumber\\
+\frac{c_0 z(0)}{(EL)^3}+\frac{c_1 z(1)}{EL}+c_2 z(2)(EL)+c_3 z(3)(EL)^3
\, .\nonumber
\end{eqnarray}
Notice that the term $\partial A_\infty / \partial W $ is not present 
because  in this case $A_\infty$ is a constant.
The $a_1$ term would be the quenched energy shift, but it now depends on the 
 choice of operator $S$ and receives contributions from $\eta'$ double-poles. 
The remaining $a_i$'s come from the $\eta'$ double-pole insertions 
giving an anomalous behaviour in time. 
The terms proportional to the $b_i$'s depend on the sink and are
different from the corresponding ones in full QCD. A consequence
of the lack of unitarity is that they cannot be eliminated using
the procedure described in sec.~\ref{sec:cf}. 
The terms proportional to the $c_i$'s correspond to the finite-volume 
corrections. They are no longer universal and the last two-terms, 
which come from the $\eta^\prime$ double pole, appear prima-facie 
to diverge in the large-volume limit. We now discuss this point further.

\section{VOLUME DEPENDENCE}
The apparent volume dependence in eq.~(\ref{scalarff}) 
is rather strange. By evaluating
the matrix element in infinite volume and euclidean metric no divergences 
appear away from the chiral limit, and one would hence expect that the 
large volume limit of the finite volume matrix element should exist.
In spite of appearances this could be the case, since at fixed
physical momenta $\vec q$ (so that the exitation level $n$ 
grows as the volume is increased), 
the $z(i)$'s scale with the volume, for instance  
$z(0)=-\nu\propto L^2$~(\footnote{where $\nu=\sum_{\Omega_{\vec q}}$ 
with $\Omega_{\vec q}=\{\vec{n}\in Z^3:|2\pi\vec{n}/L|=|\vec{q}|\}$}).
By dimensional analysis and numerical evaluation, it is possible to verify 
that the $z(i)$'s decrease with $L$ fast enough to compensate the 
powers of $L$ appearing in eq.~(\ref{scalarff}). 
However we do not yet have a proof of this 
since it is not possible to perform the large volume limit 
at fixed physical energy without the quantization condition.

\section{CONCLUSIONS}
We have shown that quenching artifacts in $K\to\pi\pi$, $\Delta I=1/2$ matrix
elements are very important and might invalidate the usual procedure 
for extracting physical matrix elements from lattice CF. 
Problems related to the non-unitarity of the quenched theory also affect 
$K\to\pi\pi$ matrix elements with $\Delta I=3/2$, but they do not appear at 
one loop in $\chi PT$ and are therefore less severe.

The problems we have raised do not necessarily have a solution within the quenched 
approximation, which, it must be remembered, is an intrinsically unphysical theory.
However, it is worth exploring whether there is some practical 
way forward, which is not obviously inconsistent.

Quenched QCD is perturbatively reproduced in the $\chi$PT framework 
by enlarging the set of particle states to contain new unphysical mesons.
This extension can be realized in two different, but perturbatively equivalent,
ways either with the supersymmetric enlargement of the chiral group~
\cite{92Bernard} or with the replica method~\cite{00Damgaard}.
We are currently investigating properties of the enlarged set of states
and the behaviour of weak matrix elements between those states, including 
the unphysical ones~\cite{02scalar}.
In particular, we are exploring whether it might be possible  
that some of those universal relations 
of full QCD mentioned in sec.~\ref{sec:intro} can be generalized to the 
quenched approximation, once the extended
set of external states is considered.

\end{document}